\documentclass[aps,prl,a4paper, reprint]{revtex4-1}

\usepackage{placeins}
\usepackage{graphicx}
\usepackage{dcolumn}
\usepackage{bm}
\usepackage[]{placeins}
\usepackage{subfigure}
\usepackage{amsmath}

\newcommand{\LJs}{BMLJ65~}
\newcommand{\LJl}{BMLJ1560~}

\newcommand{\av}[1]{{\langle #1 \rangle}}

\begin{document}
\title{Anomalous diffusion of driven particles in supercooled liquids}
\date{\today}
\author{Carsten F. E. Schroer}
\email{c.schroer@uni-muenster.de}
\author{Andreas Heuer}
\email{andheuer@uni-muenster.de}
\affiliation{Westf\"alische Wilhelms-Universit\"at M\"unster, Institut f\"ur physiklaische Chemie, Corrensstra\ss e 28/30, 48149 M\"unster, Germany}
\affiliation{NRW Graduate School of Chemistry, Wilhelm-Klemm-Stra\ss e 10, 48149 M\"unster, Germany}

\begin{abstract}
 We have performed non-equilibrium dynamics simulations of a binary Lennard-Jones mixture in which an external force is applied on a single tagged particle. For the diffusive properties of this particle parallel to the force superdiffusive behavior at intermediate times as well as giant long-time diffusivity is observed. A quantitative description of this non-trivial behavior is given by a continuous time random walk analysis of the system in configuration space. We further demonstrate, that the same physical properties which are responsible for the superdiffusivity in non-equilibrium systems also determine the non-Gaussian parameter in equilibrium systems.
\end{abstract}

\maketitle

%1.)Motivation: Horbach-Superdiff. bekannt. Bisherige Theorie: 1) Bouchaud -=> Nicht-Stationär (Problem) 2) Jack et al für kinetisches Spinmodell Info über erhöhte Langzeitdiffusion. Im selben Modell wurde allerdings auch eine negative Nichtlinearität vorhersagt für die Mobilität => Frage nach Anwendbarkeit dieser Ideen? Ziel des papers nennen:

%Dynamical heterogeneities are one of the main characteristic of supercooled liquids. The existence of slow and mobile regions results in equilibrium system into interesting properties like non-exponential relaxation and violation of Stokes-Einstein relation \cite{Binder2005}.
\emph{Introduction.} Due to the distinct multi-particle dynamics of glass-forming systems several interesting properties can be observed like the occurrence of dynamical heterogeneities \cite{Ediger2012} or the violation of the Stokes-Einstein relation \cite{Fujara1992,Bertier2004}. In the non-equilibrium situation, the observed phenomena can become even more complex. Recently, Winter et al. performed computer simulations of a tracer particle which is driven by a constant external field trough a binary Yukawa fluid \cite{Winter2012}. It was shown that for this microrheological simulation the diffusive properties of the particle become highly anisotropic: While the mean squared displacement (MSD) of the tracer particle perpendicular to the force direction $\av{x_\perp^2}(t)$ increases with increasing force but still displays a diffusive behavior, the centered MSD parallel to the force direction
\begin{equation}
 \sigma^2(t)= \av{x_\parallel^2(t)}-\av{x_\parallel(t)}^2
\end{equation}

displayed a superdiffusive behavior at the observed time range. This result was rationalized in terms of a special type of biased trap model \cite{Bouchaud1990} in which a superdiffusive behavior is predicted due to rising fluctuations. Therefore, it was stated that the diffusion constant for the parallel direction of the tracer particle does not exist \cite{Winter2012}. However, this model has to be regarded with care because the rising fluctuations would lead to a permanently increasing energy barriers. This scenario is difficult to reconcile with the observed stationary behavior. Mode-coupling theory has been very successful to predict, e.g., the nonlinear mobility dependence on the applied force in microrheological simulations \cite{Gazuz2009, Gnann2011,Harrer2012a, Harrer2012}. Interestingly, this superdiffusive behavior could not be reproduced \cite{Harrer2012}.

A different approached is used by Jack et al. \cite{Jack2008}. Motivated by the analysis of an one-dimensional spin facilitated models, they performed an analytical calculation for a biased continuous time random walk (CTRW). For this model, a diffusive regime is predicted for long times. This diffusive regime is characterized by a strong dependence on the width of the used waiting time distribution. Broader waiting time distributions lead to a dramatic increase of spatial fluctuations, denoted as "giant diffusivity" \cite{Lee2006}.

The key goal of this paper is to elucidate the properties of the superdiffusivity in the driven particle dynamics. First, we present a formal expression which relates the superdiffusivity to dynamic heterogeneities in the CTRW framework. Second, for the trajectories of a glass-forming model system we can extract the relevant observables from an appropriate CTRW analysis and predict the superdiffusive behavior in a quantitative way. For the long-time limit our expression reduces to the giant diffusivity as calculated in Ref.\cite{Jack2008}. Third, we are able to show that the superdiffusivity has a deep physical connection to the non-Gaussian parameter (NGP) in equilibrium, thereby establishing a strong connection between the non-equilibrium and the equilibrium dynamics of glass-forming systems.

%2) Saubere Abbildung auf CTRW mit MB-Eigenschaften (Zusammenfassung paper I)

\emph{Simulations.} We have performed computer simulations of a binary mixture of Lennard-Jones particles \cite{Kob1995} (BMLJ) which we have extended by applying a constant force on one randomly selected particle. Constant temperature conditions are ensured by using a Nos\'e-Hoover thermostat~\cite{Nose1984}. By applying a suitable minimization procedure it is possible to track minima of the PEL, called inherent structures (IS), which the system had explored during its time evolution. Analogous to equilibrium simulations \cite{DoliwaHopping,Rubner2008, Heuer2008} we have recently demonstrated \cite{Schroer2012}, that the time evolution of a small stationary non-equilibrium system (consisting of $65$ particles, denoted as BMLJ65) can be analogous to equilibrium systems \cite{DoliwaHopping, Rubner2008} described in terms of a continuous time random walk (CTRW) of the system between coarse grained minima, called metabasins (MB). This projection allows for a discretization of the system trajectory into dynamical events, which are characterized by the distribution of particle displacements during one transition, and a broad waiting time distribution. As shown in \cite{Schroer2012} the linear and nonlinear response only shows very small finite size effects. Here we will also show that the results of this work can be transferred to the properties of large systems as well.

\begin{figure}
 \includegraphics[width=0.40\textwidth]{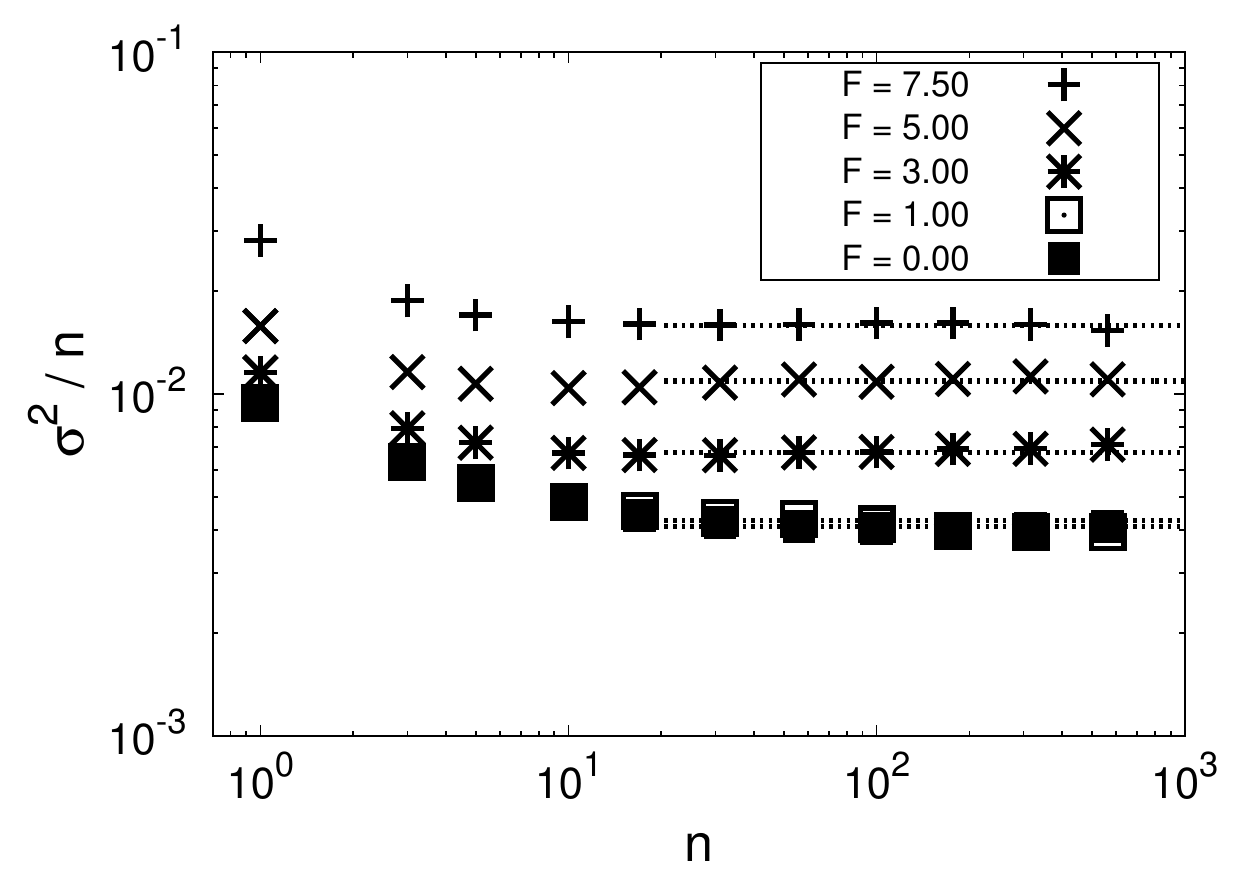}
 \caption{Centered MSD $\sigma^2(n)$ of \LJs as a function of the number of transitions $n$ at a temperature $T=0.475$. The dashed lines indicate the diffusive lengths $a^2_\parallel$ parallel to the force direction.}
 \label{figSigma2n}
\end{figure}

\begin{figure}
 \includegraphics[width=0.40\textwidth]{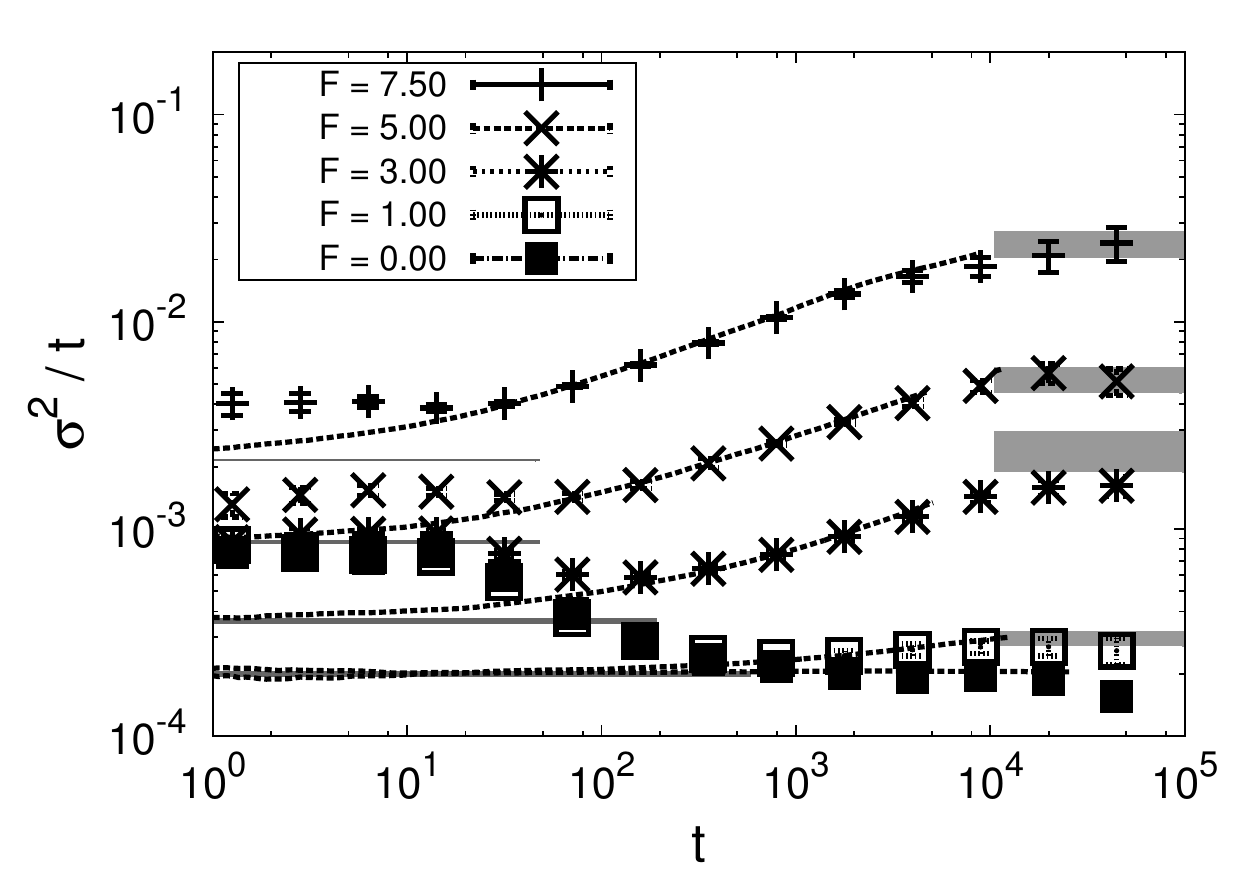}
 \caption{Centered MSD $\sigma^2(t)$ of \LJl as a function of time $t$ at a temperature $T=0.475$. The dashed lines correspond to the theoretical prediction by Eq.~\ref{eqSigma2}. The gray bars at short times indicate the equilibrium-like diffusion (Eq.~\ref{eqDiff}), at long times the long-time diffusion (Eq.~\ref{eqGiantDiff}) under consideration of the numerical error.}
 \label{figSigma2}
\end{figure}

\emph{Results.} Focusing on the diffusive behavior of the tracer particle parallel to the force direction, our approach offers two different routes to define the centered MSD: On the one hand, one can consider the centered MSD after a certain number of MB transitions $n$ on the other hand, it can be evaluated after a certain time $t$. In the following we will distinguish between these quantities by writing $\sigma^2(n)$ and $\sigma^2(t)$, respectively. Similar to the equilibrium dynamics \cite{DoliwaHopping}, $\sigma^2(n)$ grows linearly after more than $\sim20$ transitions (see Fig.~\ref{figSigma2n}). In marked contrast, $\sigma^2(t)$ displays a superdiffusive behavior (see Fig.~\ref{figSigma2}) as it was reported for the binary Yukawa fluid~\cite{Winter2012}. From $\sigma^2(n)$ one can define for large $n$ the diffusive length scale $a^2_\parallel$ via
\begin{equation}
 \label{eqDefa2}
 a^2_\parallel = \lim_{n\to\infty} \frac{\sigma^2(n)}{n} \mathrm{;}
\end{equation}
see also \cite{Schroer2012}.

 To achieve a more quantitative understanding and to unravel the surprising qualitative differences between $\sigma^2(n)$ and $\sigma^2(t)$, we have performed an analytical calculation of $\sigma^2(t)$ within the CTRW framework. We start with a one-dimensional CTRW with an elementary step $x_{i,\parallel} = a_{i,\parallel} + \Delta x_\parallel$. $\Delta x_\parallel$ denotes the average translation the particle performs along the force direction during one MB transition. As shown in \cite{Schroer2012}, $\Delta x_\parallel$ is basically proportional to $F$ in the whole force interval considered in this work. $a_{i,\parallel}$ is considered to be the remaining translational length with $\av{a_{\parallel}}=0$. Successive steps are regarded as uncorrelated so that $\av{a_{i,\parallel}a_{j,\parallel}} = \delta_{i,j}~\av{a^2_\parallel}$. For reasons of consistency with our previous work, we will further denote $\av{a^2_\parallel}$ simply as $a^2_\parallel$. Then, the MSD of the particle is given by the sum over all steps $n$ which were performed up to a time~$t$:
\begin{equation}
\av{x^2_\parallel(t)}=\av{(\sum_{i=0}^{n(t)}{(a_{i,\parallel} + \Delta x_\parallel)})^2} .
\end{equation}
 
%$$=\av{\sum_{i=0}^{n(t)}a_{i,\parallel}^2+\sum_{i=0}^{n(t)}a_{i,\parallel}a_{j,\parallel}+\Delta x\sum_{i=0}^{n(t)}a_{i,\parallel}+ \Delta x^2_\parallel n^2(t)}$$

%$$=\av{\sum_{i=0}^{n(t)}~a_{i,\parallel}^2} + \av{\Delta x_\parallel^2 n^2}$$

This yields for the time evolution of the MSD
\begin{equation}
\label{eqMSD}
 \av{x^2_\parallel(t)}=\av{n(t)}a^2_\parallel + \av{n^2(t)}\Delta x_\parallel^2~.
\end{equation}

In Eq.~\ref{eqMSD} $\av{n}$ denotes the average number of jumps the particle performs in a certain time $t$ and $\av{n^2}$ the fluctuation of these, respectively. Considering $\sigma^2(t)$ instead of the MSD one has to subtract the squared first moment of the particle displacement which is given by$ \av{x_\parallel(t)}^2 = \av{n(t)}^2~\Delta x^2_\parallel$.

After subtracting this expression from Eq.~\ref{eqMSD} one finally obtains
\begin{equation}
\label{eqSigma2}
 \sigma^2(t) = \av{n(t)}~a^2_\parallel + [\av{n^2(t)}-\av{n(t)}^2]~\Delta x^2_\parallel~.
\end{equation}

The first term of Eq.~\ref{eqSigma2} can be identified with the one-dimensional equilibrium-like diffusive process
\begin{equation}
\label{eqDiff}
 2~D_\parallel~t = a^2_\parallel\frac{t}{\av{\tau}} .
\end{equation}

We use the term "equilibrium-like" because, as we have shown for our numerical data in~\cite{Schroer2012}, both $a^2_\parallel$ and $\av{\tau}$ display a force dependence so that the actual value of $D_\parallel$ increases with increasing force. The superdiffusivity of $\sigma^2(t)$ can be related to the latter term which is only visible in driven systems. The expression $[\av{n^2(t)}-\av{n(t)}^2]$ describes the heterogeneity of the number of occurring jumps in a certain time interval and can be directly obtained from our trajectories.

For \LJs one is able to perform a complete CTRW analysis so that each observable in Eq.~\ref{eqSigma2} is directly accessible. As it can be seen in Fig.~\ref{figSigma2} this ansatz allows to quantitatively reproduce the behavior of $\sigma^2(t)$. 

The time evolution can be divided into three regimes: At short times one observes a decay of $\sigma^2(t)$ due to local caging until a short diffusive regime is reached at $t\approx\av{\tau}$ which reflects the equilibrium-like diffusion process (Eq.~\ref{eqDiff}). The initial decay of $\sigma^2(t)/t$ can be qualitatively understood by considering, that $\sigma^2(n)/n$ displays a similar decay due to slightly forward-backward correlations until the constant diffusive length $a^2_\parallel$ is reached at $n\approx20$ transitions (see~\cite{DoliwaHopping} for further details). It is important to notice that only in case of small forces the minimum of $\sigma^2(t)/t$ indicates the true value of $D_\parallel$ while at higher forces it is already superimposed by superdiffusive contributions. At intermediate times, one observes a superdiffusive behavior which is caused by the nonlinear evolution of $[\av{n^2(t)}-\av{n(t)}^2]$. At long times, when $t$ is larger than the largest measured waiting time, one indeed observes an indication, that the MSD becomes diffusive again but with a significantly larger diffusion constant. For this particular long time behavior one is able to yield an analytical prediction by the present CTRW ansatz:

The waiting time distribution $\varphi(\tau)$ of the CTRW can be characterized by its average value $\av{\tau}$ and its variance $V =\av{\tau^2}-\av{\tau^2}$. Due to the central limit theorem the distribution of the cumulated waiting time $\tau_n$ of a large number of jumps $n$, $P_n(\tau_n)$, is given by
\begin{equation}
\label{eqCentralLimT}
\lim_{n\rightarrow\infty} P_n(\tau_n) \propto \exp{\left(\frac{(\tau_n-n\av{\tau})^2}{2Vn}\right)} .
\end{equation}

The corresponding probability to find $n$ jumps in a large time interval $t$, $P_t(n)$, is directly related to $P_n(\tau_n)$. With the substitution $n = t/\av{\tau}$ and identifying $\tau_n = t$ we obtain from Eq.~\ref{eqCentralLimT} the expression
\begin{equation}
\label{eqPnt}
\lim_{t\rightarrow\infty} P_t(n) \propto \exp{\left(\frac{(n-\frac{t}{\av{\tau}})^2}{2V\frac{t}{\av{\tau}^3}}\right)} .
\end{equation}

Determination of the second moment of $P_t(n)$ yields
\begin{equation}
\label{eqLongtimeHeterogeneity}
\lim_{t\rightarrow\infty} [\av{n^2(t)}-\av{n(t)}^2] \frac{1}{t} = \frac{V}{\av{\tau}^3} = \left[\frac{\av{\tau^2}}{\av{\tau}^2}-1\right]\frac{1}{\av{\tau}}
\end{equation}
and hence, by combining Eq.~\ref{eqLongtimeHeterogeneity} and Eq.~\ref{eqSigma2}, for the long-time behavior of $\sigma^2(t)$
\begin{equation}
\label{eqGiantDiff}
\lim_{t\rightarrow\infty} \frac{\sigma^2}{t} = D_\parallel f = D_\parallel \left[1+\frac{\Delta x_\parallel^2}{a^2_\parallel}\left(\frac{\av{\tau^2}}{\av{\tau}^2}-1\right)\right] .
\end{equation}

In this equation, $f$ describes the factor which relates the equilibrium-like and the long-time diffusion constants. Independent from our derivation Jack et al. analytically obtained a similar result for the giant diffusivity by considering the Montroll-Weiss equation of a biased CTRW~\cite{Jack2008}. The long-time diffusion constant $D_\parallel f$ was already indicated in Fig.~\ref{figSigma2}. For BMLJ65, it is possible to explicitly compute the long-time diffusivity because of the direct access to the CTRW observables in Eq.~\ref{eqGiantDiff}. 
Importantly, it is also possible to estimate the long-time behavior for larger system sizes. As shown in the supplementary material the numerically observed degree of superdiffusivity is fully compatible with the theoretical expectation.

The presented ansatz allows one to give an explicit criterion how long a particle requires to reach the diffusive regime. It is related to the applicability of the central limit theorem and thus to the width of the waiting time distribution: The narrower the waiting time distribution, the earlier the particle becomes diffusive. Since the application of a strong microrhological perturbation causes a narrowing of the waiting time distribution \cite{Schroer2012} one expects an earlier advent of the long-time diffusivity at high forces. This behavior can be qualitatively observed in Fig.~\ref{figSigma2} as well.

Besides the MSD of a driven particle, the heterogeneity of MB transitions can also be observed in equilibrium systems by analyzing the NGP $\alpha_2(t)$ of the one-dimensional particle displacement which is defined as
\begin{equation}
 \alpha_2(t) = \frac{\av{x^4(t)}-3\av{x^2(t)}^2}{3\av{x^2(t)}^2}.
\end{equation}

Using the ansatz of an unbiased CTRW one obtains for the NGP
\begin{figure}
 \includegraphics[width=0.40\textwidth]{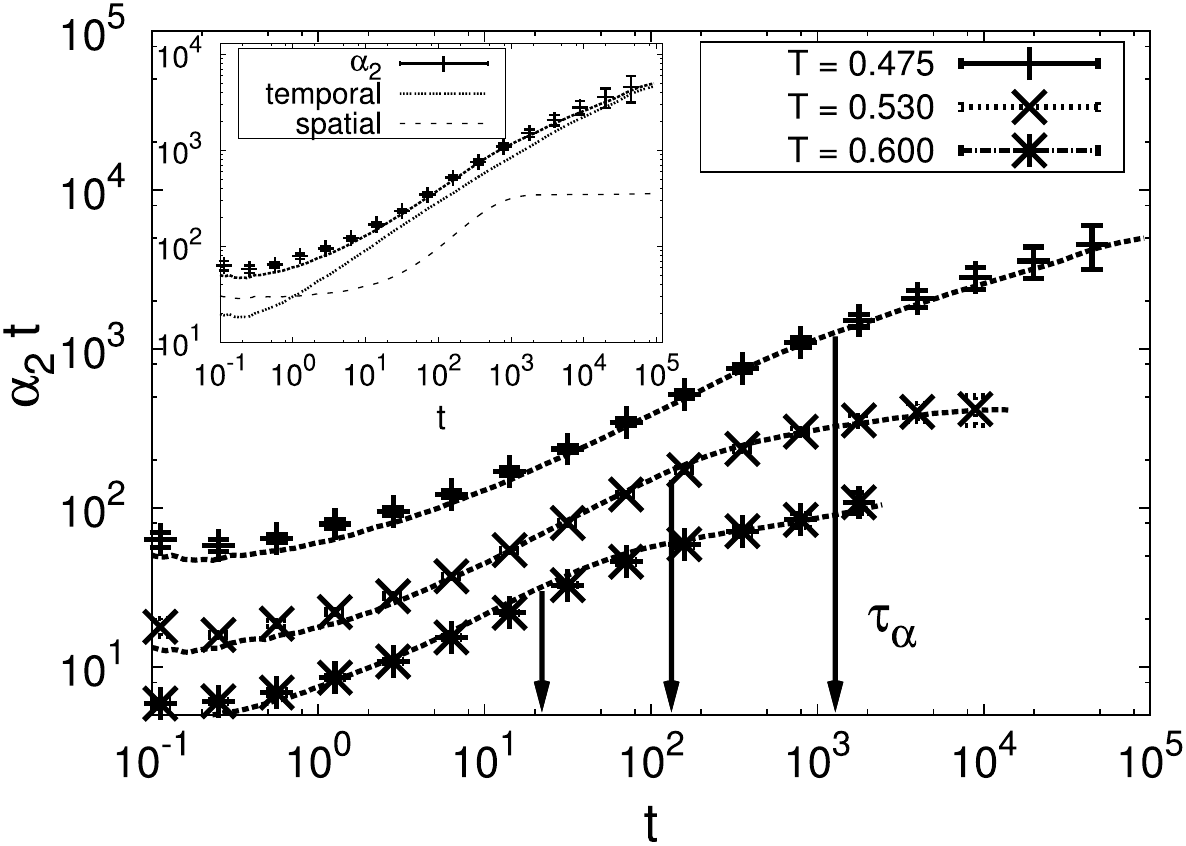}
 \caption{One-dimensional NGP $\alpha_2(t)$ multiplied by time $t$ as a function of $t$. The dashed lines corresponds to the theoretical prediction in Eq.~\ref{eqAlpha2}, the arrows indicate the structural relaxation time $\tau_\alpha$ for the different temperatures. Inset: $\alpha_2(t)~t$ at a temperature $T=0.475$ together with its temporal and spatial contributions (see text).}
 \label{figAlpha2}
\end{figure} 
\begin{align}
 \label{eqAlpha2}
 \alpha_2(t) = & \frac{[\av{n^2(t)}-\av{n(t)}^2]}{\av{n(t)}^2}+ \int{dn~P_t(n)A(n)} \mathrm{.}
\end{align}

The latter term describes the non-Gaussianity of the cumulated displacements $a_n$ after $n$ transitions
\begin{equation}
A(n) = \frac{[\av{a^4_n}-3\av{a^2_n}^2]}{3\av{a^2_n}^2}
\end{equation}
weighted by the probability to find exactly $n$ transitions at a time $t$. In what follows we use the approximation that $\int{dn~P_t(n)A(n)}\approx A(\av{n(t)})$.

Eq.~\ref{eqAlpha2} contains two contributions: The first term includes the heterogeneity of the performed jumps $n$ in a certain time interval. It is the same quantity which was observed to be responsible for the superdiffusive behavior in the non-equilibrium system. Because this term is independent of any length scales, one can regard it as a measure for the temporal heterogeneities of the system dynamics. The latter term reflects spatial heterogeneities of the elementary MB transition which become less important at a larger number of transitions because the distribution of the cumulated lengths approaches a Gaussian shape.

In Fig.~\ref{figAlpha2} $\alpha_2(t)\cdot t$ is shown at different temperatures together with the theoretical prediction by Eq.~\ref{eqAlpha2}. Please note that, as it was also shown by Liao et al.~\cite{Liao2001}, in case of transitions between adjacent MB, $\alpha_2(t)$ displays a monotonic decay. This behavior can be understood by considering that the initial growth of $\alpha_2(t)$ is caused by vibrational parts (short times) and the $\beta$-relaxation process (at intermediate times)~\cite{Kob1995,Caprion2000} while, by construction, MB trajectories only resolve the $\alpha$-relaxation process~\cite{Rehwald2012}. As one can see, the theoretical prediction allows one to fully describe the NGP at each temperature. One further observes that for long times, $\alpha_2(t)\cdot t$ approaches a constant which corresponds to a decay of $\alpha_2(t)\propto1/t$ which is exactly the expectation for $[\av{n^2(t)}-\av{n(t)}^2]/\av{n(t)}^2$ when the central limit theorem becomes valid (see Eq.~\ref{eqLongtimeHeterogeneity}).

In the inset of Fig.~\ref{figAlpha2} the different temporal and spatial contributions to $\alpha_2(t)\cdot t$ are shown. At very short times, the behavior of $\alpha_2(t)\cdot t$ is mainly determined by $A(\av{n(t)})$ while above $t\approx 10^1$, the temporal part of Eq.~\ref{eqAlpha2} is found to be the major contribution. At $t\approx 10^4$, one observes $A(\av{n(t)})\propto 1/t$ which indicates that the central limit theorem starts to hold for the distribution of the spatial displacement. Indeed, for the waiting time distribution the central limit theorem is only fulfilled on a larger time scale, so that there is still a growth of $\alpha_2(t)\cdot t$.

%One can observe that the spatial contribution approaches a constant at a time scale of the structural relaxation time $\tau_\alpha$, i.e., at $\tau_\alpha$ the central limit theorem of the elementary step is fulfilled. However, the temporal part still displays a growth at $\tau_\alpha$ so that it the central limit theorem of the waiting time distribution is fulfilled at much longer times. Interestingly, this is the major contribution to the long-time behavior of $\alpha_2(t)$.

%

It is know from a comparison between mode-coupling theory and Brownian dynamics simulations of BMLJ~\cite{Flenner2005}, that mode-coupling theory tends to strongly underestimate the magnitude of $\alpha_2(t)$ in the diffusive regime. This differences between theory and simulation are also known for the hard-sphere system \cite{Fuchs1998,Voigtmann2004}. Recently, it was further reported, that simplified mode-coupling theory models cannot reproduce the superdiffusive behavior of a driven particle along its force direction~\cite{Harrer2012}. Therefore, it is quite remarkable that the CTRW approach enables to relate both, the non-Gaussianity of the equilibrium system and the superdiffusive behavior of the stationary non-equilibrium system to have the same origin, reflecting the presence of the dynamic heterogeneities. One thus might argue whether mode-coupling theory, possibly due to its dependence on average quantities \cite{Fuchs1998}, is not able to fully describe these fluctuations. As a consequence \emph{both} effects cannot be qualitatively reproduced.

\emph{Summary.}
In the present paper we have demonstrated, that a model of a biased CTRW allows to fully predict the anomalous diffusion of a driven particle in a supercooled medium which is characterized by equilibrium-like diffusion, superdiffusivity and long-time diffusivity. It was further shown, that the origin of the superdiffusivity results from temporal fluctuations of the system dynamics which become visible due the applied bias. Indeed, also in equilibrium the same fluctuations are present and determine the evolution of the NGP $\alpha_2(t)$. Therefore, the connection between superdiffusive behavior and non-Gaussianity is a remarkable example, how non-equilibrium dynamics also enables a deeper physical understanding of the equilibrium system, by uncovering essential underlying physical properties.

This work was supported by DFG Research Unit 1394 "Nonlinear Response to Probe Vitrification". Furthermore, C. F. E. Schroer thanks the NRW Graduate School of Chemistry for funding. We acknowledge helpful discussions with J. Horbach, C. Rehwald and D. Winter.
%analyzed the diffusive behavior of a single driven particle in a supercooled liquid along the force direction. On the basis of a simple toy model it was demonstrated that a biased CTRW allows to explain both, superdiffusive behavior as well as long-time diffusivity along the force direction. We have proven this results by considering numerical results of a non-equilibrium molecular dynamics simulations which was analyzed in term of a CTRW in configuration space. 

%The origin of anomalous was identified as temporal fluctuations of the system's dynamics which become visible due to the application of the bias. However, this fluctuations were also present in equilibrium system which we have demonstrated by analyzing the non-Gaussian parameter in terms of a non-biased CTRW.

\title{Supporting material}
%\date{\today}
%\author{Carsten F. E. Schroer}
%\email{c.schroer@uni-muenster.de}
%\author{Andreas Heuer}
%\email{andheu@uni-muenster.de}
%\affiliation{Westf\"alische Wilhelms-Universit\"at M\"unster, Institut f\"ur physiklaische Chemie, Corrensstra\ss e 28/30, 48149 M\"unster, Germany}
%\affiliation{NRW Graduate School of Chemistry, Wilhelm-Klemm-Stra\ss e 10, 48149 M\"unster, Germany}

% \begin{abstract}
%  We have performed non-equilibrium dynamics simulations of a binary Lennard-Jones mixture in which an external force was applied on a single tagged particle. For the diffusive properties of this particle parallel to the force superdiffusive behavior at intermediate times as well as long-time diffusivity is observed. A qualitative description of this non-trivial behavior is given by a continuous time random walk (CTRW) analysis of the system in configuration space. We further demonstrate, that the same physical properties which are responsible for the superdiffusivity are present in equilibrium systems as well and determine the non-Gaussian parameter.
% \end{abstract}

%\maketitle
\newpage
\section*{Long-time diffusivity of larger systems}

As we had demonstrated, the continuous time random walk (CTRW) analysis allows one to predict the long-time diffusivity of the driven particle as

\begin{equation}
\label{eqfDCTRW}
 \lim_{t\to\infty} \frac{\sigma^2(t)}{t} = D_\parallel~f=D_\parallel~\left[1+\frac{\Delta x_\parallel^2}{a^2_\parallel}\left(\frac{\av{\tau^2}}{\av{\tau}^2}-1\right)\right]~.
\end{equation}

For a binary mixture of $N=65$ Lennard-Jones particles (BMLJ65) the required CTRW quantities can be explicitly computed from the metabasin (MB) trajectories. For systems larger then \LJs it is not possible because there is no direct access to the required quantities. Unfortunately it is known, that especially the structural relaxation time $\tau_\alpha$, which reflects the higher moments of the waiting time distribution, displays significant finite-size effects \cite{Heuer2008,Rehwald2010}. Therefore, one must expect, that the degree of superdiffusivity will be different at larger system sizes. It is nevertheless possible to estimate the magnitude of long-time diffusivity from the real space trajectories of larger systems.

We start with a relation between the average waiting time $\av{\tau}$ and the diffusion constants $D_{\parallel/\perp}$ of the tracer particle parallel and perpendicular to the direction of the force. It is given by

\begin{equation}
\label{eqD}
 D_{\parallel/\perp} = \frac{a^2_{\parallel/\perp}}{\av{\tau}}
\end{equation}

where $a^2_{\parallel/\perp}$ denote the apparent diffusive lengths during one MB transition \cite{DoliwaHopping, Schroer2012}. Furthermore, the second moment of the waiting time distribution $\av{\tau^2}$ is connected to $\tau_\alpha$ by~\cite{Rubner2008, Rehwald2010}

\begin{equation}
\label{eqTauAlpha}
 \lim_{q\to\infty} \tau_{\alpha}(q) = \tau_{\alpha} = \frac{\av{\tau^2}}{2\av{\tau}} .
\end{equation}

Please note, that in a strict sense the relation between $\lim_{q\to\infty}\tau_\alpha(q)$ and $\av{\tau^2}$ is only valid for BMLJ65. We will identify the relevant value of $\tau_\alpha$ further below. Inserting Eq.~\ref{eqD} and Eq.~\ref{eqTauAlpha} in Eq.~\ref{eqfDCTRW} yields

\begin{equation}
\label{eqfDReal}
 D_\parallel f = \frac{a^2_\parallel}{a^2_\perp}D_\perp \left[ \frac{\Delta x^2_\parallel}{a^2_\parallel}\left(\frac{4~\tau_\alpha~D_\perp}{a^2_\perp}-1\right)\right] \mathrm{.}
\end{equation}

In contrast to Eq.~\ref{eqfDCTRW}, Eq.~\ref{eqfDReal} only depends on microscopic lengths and quantities which are measurable in real space. Please note, that the diffusion constant $D_\parallel$ was substituted by $\frac{a^2_\parallel}{a^2_\perp}D_\perp$ because $D_\parallel$ is not accessible due to the superdiffusive behavior of~$\sigma(t)$.

\begin{figure}
 \includegraphics[width=0.45\textwidth]{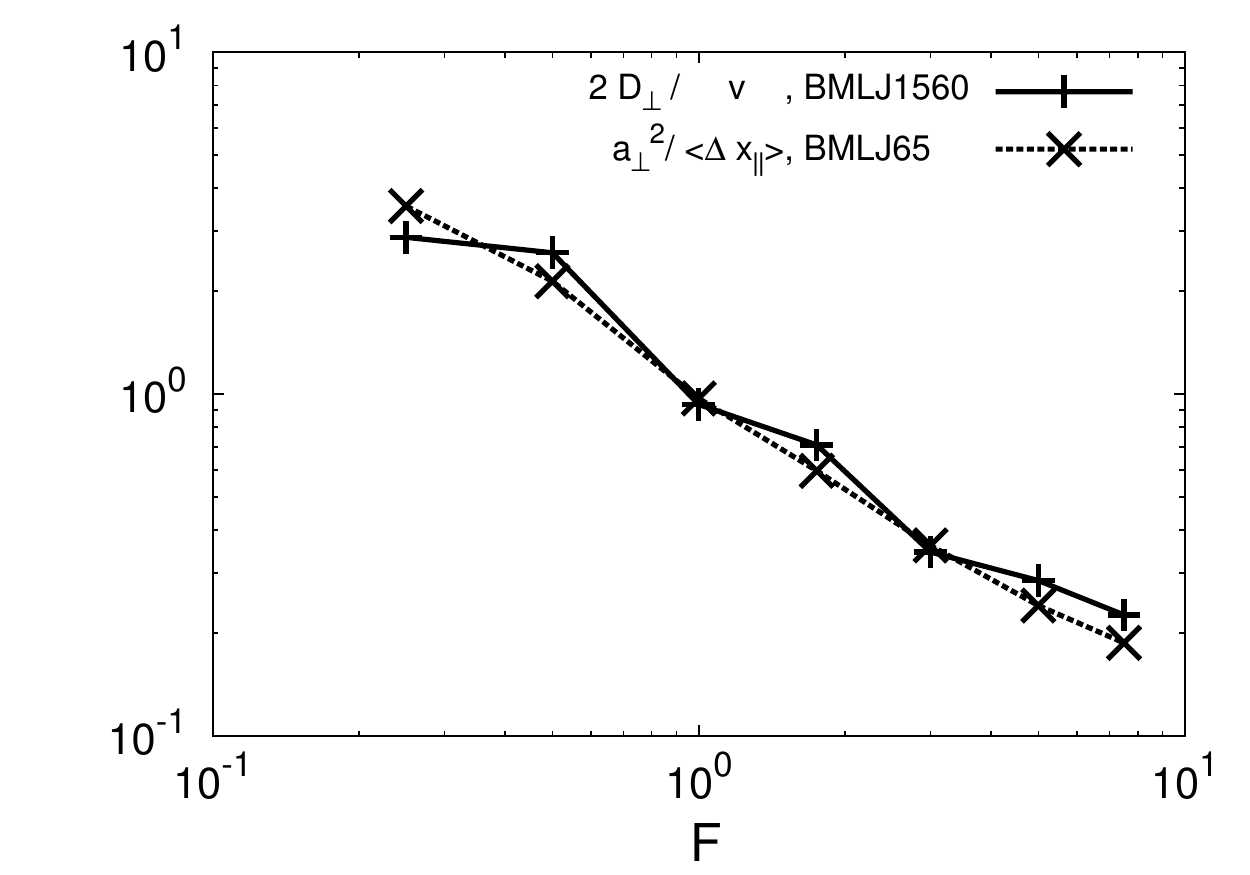}
 \caption{Ratios of the diffusion coefficient $D_\perp$ perpendicular to the force direction and the drift velocity $v$ of \LJl and the corresponding microscopic length scales $a^2_\perp$ and $\Delta x_\parallel$ of \LJs versus the applied Force $F$. Both systems were simulated at $T=0.475$.}
 \label{figScaling}
\end{figure}

We now assume that the microscopic length scales in Eq.~\ref{eqfDReal} are the same for \LJs and larger systems. This assumption is made because the microscopic lengths refer to the local spatial displacement of single particles during single relaxation processes which should be largely independent from the system size. To verify this, we have computed the ratio of the drift velocity $v$ and the perpendicular diffusion constant $D_\perp$ of a system with $N=1560$ particles (BMLJ1560). Because $v=\frac{\Delta x_\parallel}{\av{\tau}}$ and Eq.~\ref{eqD} (for BMLJ65), this ratio can be identified with $a^2_\perp/2 \Delta x_\parallel$. If the assumption holds, $a^2_\perp/\Delta x_\parallel$ of \LJs should be essentially the same as $2 D_\perp/v$ of BMLJ1560. The corresponding plot is shown in Fig.~\ref{figScaling}. One can observe, that the mismatch between these curve is less than 20\% so that the independence of the microscopic length scales on system size indeed seems to be reasonable.

The missing observable $\tau_\alpha$ in Eq.~\ref{eqfDReal} is determined by computing the incoherent scattering function 
\begin{equation}
 S_\perp(q,t) = \av{\cos{(q(x_\perp(t)-x_\perp(t_0))^2)}}
\end{equation}
of the tracer particle \emph{perpendicular} to the force direction. The reason for the choice of the perpendicular direction is, that one would expect, analogous to the superdiffusive behavior of $\sigma^2(t)$, an additional relaxation along the parallel direction which does not reflect the underlying waiting time distribution. Following the procedure of Rehwald et al. \cite{Rehwald2010} we have computed the apparent wave-vector relaxation 
\begin{equation}
 \tau(q) = \int_0^\infty{dt~S_\perp(q,t)} \mathrm{.}
\end{equation}
and fitted the resulting curves in the range $q\leq 8$ by
\begin{equation}
 \tau(q) = \tau_\alpha + \frac{1}{q^2\cdot D_\perp}\mathrm{,}
\end{equation}
finally allowing us to determine the local relaxation time $\tau_\alpha$.

\begin{figure}
 \includegraphics[width=0.45\textwidth]{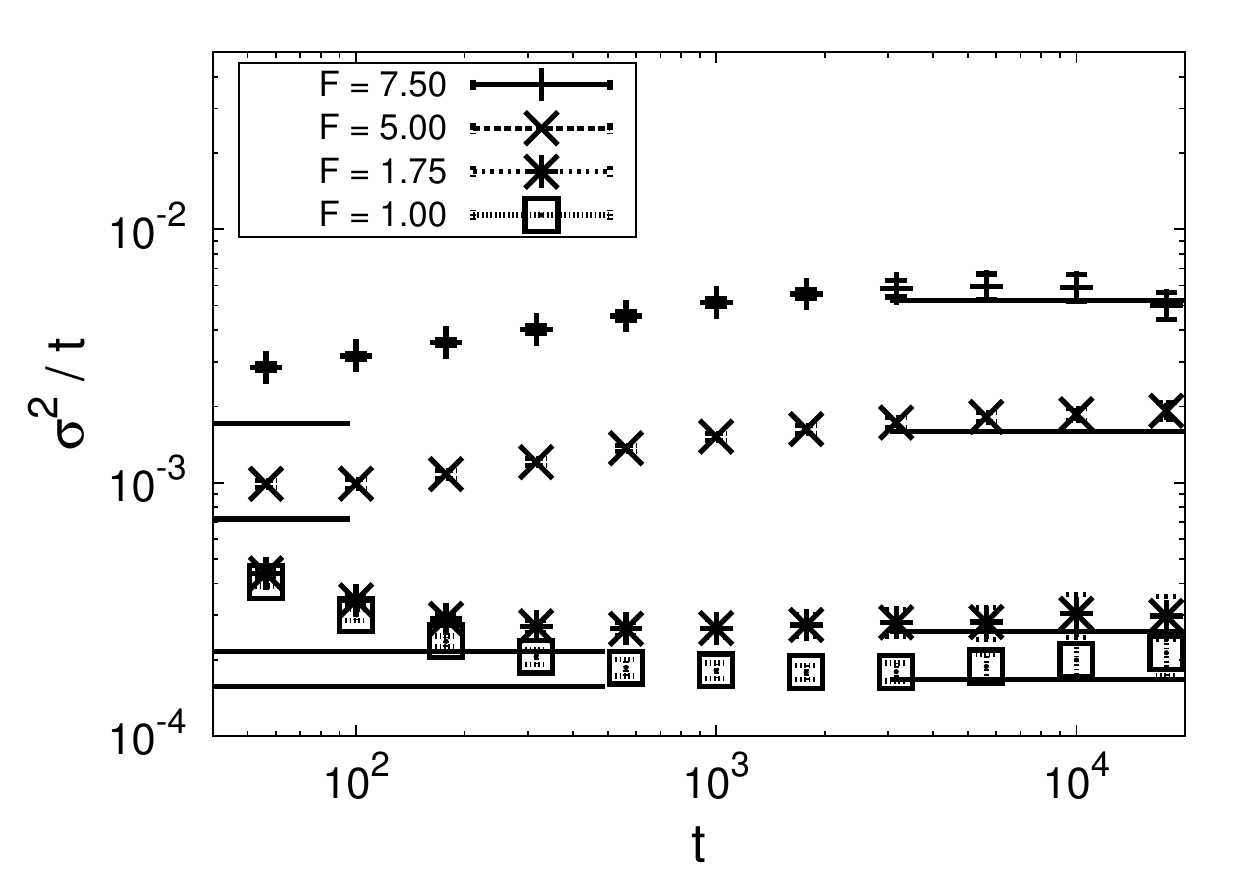}
 \caption{Centered MSD $\sigma^2(t)$ versus time $t$ for BMLJ1560 at $T=0.475$. The solid lines at short times correspond to the equilibrium-like diffusion constant (Eq.~\ref{eqD}), the solid lines at long times to the predicted long-time diffusivity (Eq.\ref{eqfDReal}).}
 \label{figSigmaLarge}
\end{figure}

In Fig.~\ref{figSigmaLarge} $\sigma^2(t)$ of \LJl is shown together with the predicted equilibrium-like and long-time diffusion constants. One observe a quantitative agreement between the numerical data and the theoretical prediction. Compared to BMLJ65, the factor between equilibrium-like and long-time diffusion is smaller and the long-time diffusivity appears earlier. This behavior indicates a narrowing of the underlying waiting time distribution of BMLJ1560. One can rationalize this by regarding the large system as a composition of small elementary system which are coupled so that relaxation processes of single subsystems induce further relaxation events in adjacent systems. This concept has been discussed in \cite{Rehwald2010, Rehwald2012}.

\end{document}